\documentclass[12pt,a4paper]{article}
\usepackage{amsmath}
\usepackage{amsxtra}
    \usepackage{amstext}
    \usepackage{amssymb}
    \usepackage{latexsym}
\usepackage{graphics}

\newcommand{\D}{{\cal D}}
\newcommand{\half}{\frac{1}{2}}

\newcommand{\nn}{\nonumber}

\newcommand{\pt}{\partial_{t}}

\newcommand{\p}{\vspace{6pt}\noindent}

\advance\textheight by \topskip
\makeatletter

\@addtoreset{equation}{section}
\def\section{\@startsection {section}{1}{\z@}{-8.5ex plus -1ex minus
 -.2ex}{3.3ex plus .2ex}{\large\bf}}
\def\subsection{\@startsection{subsection}{2}{\z@}{-3.25ex plus
 -1ex minus -.2ex}{1.5ex plus .2ex}{\bf}}
\def\subsubsection{\@startsection{subsubsection}{3}{\z@}{-3.25ex plus%
 -1ex minus -.2ex}{1.5ex plus .2ex}{\sl}}

\begin{document}
\begin{titlepage}
\vspace*{-2cm}
\begin{flushright}
\end{flushright}

\vspace{0.3cm}

\begin{center}
{\Large {\bf }} \vspace{1cm} {\Large {\bf A new class of integrable defects}}\\
\vspace{1cm} {\large  E.\ Corrigan\footnote{\noindent E-mail: {\tt
edward.corrigan@durham.ac.uk}} and
C.\ Zambon\footnote{\noindent E-mail: {\tt cristina.zambon@durham.ac.uk}} \\
\vspace{0.3cm}
{\em Department of Mathematical Sciences \\ University of Durham, Durham DH1 3LE, U.K.}} \\

\vspace{2cm} {\bf{ABSTRACT}}
\end{center}
An alternative Lagrangian definition of an integrable defect is provided and analyzed. The new approach is sufficiently broad to allow a description of defects within the Tzitz\'eica model, which was not possible in previous approaches, and may be generalizable. New, two-parameter, sine-Gordon defects are also described, which have characteristics resembling a pair of `fused' defects of a previously considered type. The relationship between these defects and B\"acklund transformations is described and a Hamiltonian description of integrable defects is proposed.

\vfill
\end{titlepage}

\section{Introduction}

It was noticed some years ago \cite{bczlandau,bcztoda} that an integrable field theory in two-dimensional space-time can accommodate discontinuities yet remain integrable. The fields on either side of a discontinuity are related to each other by a set of \lq defect' conditions, including the influence of a \lq defect' potential whose form is required by integrability. The defect conditions themselves are interesting since they are related, at least in the examples investigated so far, to B\"acklund transformations frozen at the location of the defect. It has been found, possibly owing ultimately to the latter observation, that defects can be supported within the $a_n^{(1)}$ series of affine Toda models \cite{mikhailov79, mikhailov80}, of which the sine-Gordon model is the first member. Intriguingly, and despite translation invariance being explicitly broken by the prescribed location, defect conditions compatible with integrability are determined simply by demanding that the defect itself be able to contribute consistently to ensure the whole system supports a conserved energy and momentum. The defect may be located anywhere (or even move at a constant speed \cite{bczsg05}), but the defect conditions apparently compensate for the evident lack of translation invariance. One might regard the defect as a state within the model whose presence is indicated by a set of defect conditions described by an additional term in the Lagrangian description rather than being a field excitation or smooth field configuration. Typically, an integrable defect will be purely transmitting and its effect does not depend upon its location, meaning it is essentially `topological'. At a classical level this is exemplified by the passage of a sine-Gordon soliton through a defect where the soliton will be delayed (or advanced), but might alternatively, according to circumstances, be absorbed by the defect or flipped to an anti-soliton \cite{bczsg05}. Similar types of behaviour are observed for the complex solitons of the $a_n^{(1)}$ models \cite{cz07}. At a quantum level, defects also appear to play a role though again they are purely transmitting and described by a transmission matrix that is compatible with the bulk scattering matrix. The purely transmitting aspect of the setup was to be expected from observations by Delfino, Mussardo and Simonetti \cite{Delf94}, but it is still of interest to see exactly how this transpires in detail. In the sine-Gordon case, the transmission matrix was anticipated by Konik and LeClair \cite{Konik97} but rederived and its properties explored in detail in \cite{bczsg05}; for other members of the $a_n^{(1)}$ series, the transmission matrices have been provided more recently \cite{cz09}. There are a number of related ideas and calculations, including perturbative checks of transmission factors for breathers, and an analysis of the interesting relationship between integrable boundary conditions and defects; some of these are explored in the article by Bajnok and Simon \cite{Bajnok}.

\p The sine-Gordon B\"acklund transformation was generalised to $a_n^{(1)}$ affine Toda models by Fordy and Gibbons \cite{Fordy80} and it seems surprising there appear to be no similarly explicit B\"acklund transformations for the other series of Toda models. However, that fact is at least consistent with the apparent absence of defects in most of these models, at least  of the kind previously considered \cite{cz09}. On the other hand, there are several types of B\"acklund transformation available in the literature for the Tzitz\'eica model \cite{Tzitzeica,TzitzeicaB1,TzitzeicaB2,TzitzeicaB3}\,\footnote{\ Note: the model introduced by Tzitz\'eica is the $a_2^{(2)}$ member of the affine Toda collection of field theories and is also known as the Bullough-Dodd or Zhiber-Mikhailov-Shabat equation.} and, therefore, one might suppose there should be a generalisation of the defect, at least for this model, and possibly for others.  The purpose of this article is to propose a generalisation by allowing a defect to have its own degree of freedom in a certain well-defined manner, which is just general enough to encompass the Tzitz\'eica model. This is reminiscent of  an idea of Baseilhac and Delius concerning dynamical boundaries \cite{Baseilhac} though it turns out to be rather different in practice. Applying the same idea to massive free fields and to the sine-Gordon model leads to new types of defect even there, encouraging the possibility of finding a more general framework that might be able to accommodate defects in all Toda models. It is interesting also to note that in the sine-Gordon model the new defects belong to a two-parameter family, which in a certain sense might be regarded as `bound states' of the  defects introduced in \cite{bczlandau}.

\p As mentioned above, the requirement of overall energy-momentum conservation is surprisingly powerful and will be the main technique employed, although, clearly, further checks are needed to verify integrability. On the other hand, previous experience strongly suggests the conditions following from momentum conservation in the presence of a defect are more or less equivalent to the restrictions imposed by integrability: for example, even if the bulk models on either side of the defect are not specified in advance, they will be severely restricted by insisting on momentum conservation once the defect is taken into account. So far, unlike the cases within the older framework, where the integrability is underpinned by a generalised Lax pair \cite{bczlandau, bcztoda}, no suitable Lax pair description describing the new framework exists yet, and it is necessary to provide alternative arguments. A small step in this direction is provided in Appendix A where it is demonstrated that the new defect conditions for the sinh/sine-Gordon equation are enough to ensure the existence of a conserved spin three charge. Other, indirect, evidence is provided in section 5 where the relationships between defects of different types and B\"acklund transformations are elaborated. Finally, a sketch of a Hamiltonian approach is given in section 6 within which defect conditions are regarded as constraints imposed at the location of the defect on the fields to either side of it.

\section{Generalising the framework}

\p Consider a defect located at the origin $x=0$ and let $u$ and $v$ be the fields on either side of it in the regions $x<0$ and $x>0$, respectively. Typically, a defect defined by B\"acklund conditions will have a discontinuity, in the sense that
while the conditions sewing the two fields at the origin constrain their derivatives the fields themselves are not prescribed. In other words, it is expected that the values of the fields approaching $x=0$ from their respective domains need not match and it should be expected that $u(0,t)-v(0,t)\ne 0.$
The basic idea to be explored here introduces a new variable $\lambda(t)$  associated with the defect itself. The simplest setup one might envisage does not directly associate dynamics to $\lambda$ but is linear in $\lambda_t$ having a Lagrangian description of the form:
 \begin{equation}\label{lambdalagrangian}
{\cal L}=\theta(-x){\cal L}_u +\theta(x){\cal L}_v+\delta(x)
\left(\frac{uv_t-vu_t}{2}+\lambda(u-v)_t-\lambda_t (u-v) -{\cal D}(u,v,\lambda)\right),
\end{equation}
where the Heaviside step function $\theta(x)$ and the Dirac delta function have been inserted to ensure the fields $u$, $v$ are restricted to their respected domains with the defect located at $x=0$.
In a sense,  $\lambda(t)$ plays the role of a Lagrange multiplier: if the potential were absent, integrating over $\lambda$ would require the discontinuity to be time-independent. However, because the potential also depends on $\lambda$ it has a more interesting effect. As we shall see, this is the case even if the potential is quadratic and the defect links two free massive fields. For the purposes of distinguishing the cases with and without the extra degree of freedom, defects of the original type ($\lambda\equiv 0$) will be called type I and those where $\lambda$ plays a role will be called type II.

\p The defect conditions at $x=0$ implied by \eqref{lambdalagrangian}
are:
\begin{eqnarray}\label{defectcondition}
  u_x&=&v_t-2\lambda_t-\frac{\partial{\cal D}}{\partial u} \\
   v_x&=&u_t-2\lambda_t+\frac{\partial{\cal D}}{\partial v} \\
 \label{defectcondition3} u_t&=&v_t+\frac{1}{2}\frac{\partial{\cal D}}{\partial \lambda}.
\end{eqnarray}
Then, it is not difficult to show directly that ${\cal E+D}$ is conserved, where ${\cal E}$ is the combined bulk contributions to the total energy from the fields $u$ and $v$. This was to be expected since time translation invariance has not been violated.

\p
On the other hand, as usual the contribution from the fields $u$ and $v$ to the total momentum is not conserved and the requirement of being able to construct a compensating contribution from the defect is highly constraining. Defining
\begin{equation}
{\cal P}=\int_{-\infty}^0 dx\, u_x u_t + \int_0^{\infty} dx\, v_x v_t,
\end{equation}
differentiating with respect to time, and using the bulk equations of motion, gives
\begin{equation}
\dot {\cal P}=\frac{1}{2}\left(u_t^2+u_x^2-2U(u)\right)_{x=0}-
\frac{1}{2}\left(v_t^2+v_x^2-2V(v)\right)_{x=0}.
\end{equation}
Using the defect conditions (and simplifying the notation on the understanding all field quantities are evaluated at $x=0$), the latter can be rewritten as
\begin{equation}\label{basicrelation}
-v_t\frac{\partial{\cal D}}{\partial u}-u_t\frac{\partial{\cal D}}{\partial v}+2\lambda_t
\left(\frac{\partial{\cal D}}{\partial u}+\frac{\partial{\cal D}}{\partial v}
+\frac{1}{2}\frac{\partial{\cal D}}{\partial \lambda}\right) +\frac{1}{2}\left(\left(
\frac{\partial{\cal D}}{\partial u}\right)^2-\left(
\frac{\partial{\cal D}}{\partial v}\right)^2\right)-U+V.
\end{equation}

\p For type I defects it would be natural to require the last piece (without any time-derivatives) to vanish and the first two pieces should be a total time derivative leading to equations for the potential ${\cal D}$:
\begin{equation}\label{typeIconditions}
\frac{\partial^2{\cal D}}{\partial u^2}=\frac{\partial^2{\cal D}}{\partial v^2},\quad \frac{1}{2}\left(\left(
\frac{\partial{\cal D}}{\partial u}\right)^2-\left(
\frac{\partial{\cal D}}{\partial v}\right)^2\right)=U-V.
\end{equation}
This was the setup originally considered in \cite{bczlandau}. In fact, as was recalled in the introduction, the conditions \eqref{typeIconditions} are highly constraining, effectively limiting $U,V$ (and $\D$) to the set of sine/sinh-Gordon, Liouville, massive or massless, free fields. In particular, the Tzitz\'eica equation is explicitly excluded. It is also worth recalling the well-known fact that the same selection of fields follows from insisting on the conservation of a spin three charge in the bulk (and that a careful analysis of the energy-like spin three charge is enough to provide the full set of integrable boundary conditions for the sine/sinh-Gordon model \cite{ghoshal}). The Tzitz\'eica equation does not allow the conservation of a spin three charge but is the one additional possibility  that arises if one instead examines a bulk conserved charge of spin five.

\p However, for type II defects, where $\lambda\ne0$, the condition on the part of \eqref{basicrelation} containing no explicit derivatives is weaker because it need not be zero as was assumed in \eqref{typeIconditions}. Rather, it should be equated with
$$\frac{1}{2}F(u,v,\lambda)\frac{\partial{\cal D}}{\partial \lambda}\equiv (u-v)_t\, F(u,v,\lambda),$$
for some function $F$ depending on $u,v$ and $\lambda$, but not their derivatives. In turn, this observation modifies the impact of the other terms.
Taking it into account and assuming the result is a total time derivative of $-\Omega$,  designed to be a functional of $u(0,t), v(0,t)$ and $\lambda(t)$, requires:
\begin{eqnarray}
 \nn \frac{\partial\Omega}{\partial u}&=&\frac{\partial{\cal D}}{\partial v}-F \\
\nn  \frac{\partial\Omega}{\partial v}&=&\frac{\partial{\cal D}}{\partial u}+F \\
    \,\frac{\partial\Omega}{\partial \lambda}&=& -2\left(
\frac{\partial{\cal D}}{\partial u}+\frac{\partial{\cal D}}{\partial v}+\frac{1}{2}
 \nn \frac{\partial{\cal D}}{\partial \lambda}\right),\\
\left(\frac{\partial \D}{\partial u}\right)^2-\left(\frac{\partial \D}{\partial v}\right)^2&=&2(U-V)+F\,\D_\lambda\, .
\end{eqnarray}
This set of equations entails a number of compatibility relations and to examine these
it is convenient to use new field coordinates defined at the defect location:
$$p=\frac{u(0,t)+v(0,t)}{2},\quad q=\frac{u(0,t)-v(0,t)}{2}.$$
Then, after a few manipulations the conditions become (and hereafter subscripts
will be used to denote partial derivatives):
\begin{eqnarray}
\nn\Omega_p&=&\D_p\\
\nn\Omega_q&=&-\D_q-2F\\
\Omega_\lambda&=&-\D_\lambda - 2\D_p.
\end{eqnarray}
Eliminating $\Omega$  leads to
\begin{eqnarray}
\nn \D_{pq}&=&-F_p \\
 \nn  \D_{\lambda p} &=& -\D_{pp} \\
 F_\lambda &=& -F_p,
\end{eqnarray}
and, from these it follows that:
$$\D=f+g, \quad F=-f_q,\quad \Omega=f-g,$$
where $g$ depends only on $\lambda$ and $q$, and $f$ depends on $q$ and $p-\lambda$.
Under these circumstances, the last, nonlinear, relation becomes
$$\D_p\D_q=2(U-V)+(f_\lambda +g_\lambda)\, F,$$
and this may also be rearranged and rewritten in terms of derivatives of $f$ and $g$:

\begin{equation}\label{fgrelation}\frac{1}{2}(f_q g_\lambda - f_\lambda g_q)=U-V.\end{equation}

\p
Interestingly, the left hand side of \eqref{fgrelation} is equal to the Poisson bracket of $f$ and $g$ regarded as functions of $\lambda$ and its conjugate momentum $\pi_\lambda =-(u-v)=-2q$. In terms of the defect energy and momentum, $\cal{D}$ and ${\Omega}$, the relationship \eqref{fgrelation} is
\begin{equation}
\{ {\cal{D}},\Omega\}=-2(U-V),
\end{equation}
an intriguing equation that relates the Poisson bracket of the energy and momentum contributed by the defect, which is non-zero because of the lack of translation invariance, to the potential difference across the defect.

\p Finally, it is worth noting that the equation \eqref{fgrelation} is powerful because all the dependence on $\lambda$ contained in the left hand side of the equation must cancel out; this significantly constrains not only $f$ and $g$ but also the potentials $U(u)$ and $V(v)$. As will be seen below the list of possibilities will now include the Tzitz\'eica model that had been excluded previously.

\section{Examples}

\p In this section, using natural ans\"atze, a number of possible solutions to \eqref{fgrelation} are given. Besides the Tzitz\'eica equation these solutions provide generalisations of already known integrable defects. However, it is not clear that the examples given exhaust all possible solutions to \eqref{fgrelation}.

\subsection{The sinh/sine-Gordon model}

\p For the sine-Gordon model, given the form of the potentials
 $$U(u) =e^{p+q}+e^{-p-q}\equiv e^u+e^{-u},\quad V(v)=e^{p-q}+e^{-p+q}\equiv e^v+e^{-v},$$
and bearing in mind the form of the constraint \eqref{fgrelation}, the most general ansatz for $f$ and $g$ is
\begin{equation}
f=Ae^{p-\lambda} +B e^{-p+\lambda}, \quad g=Ce^{-\lambda} +D e^{\lambda},
\end{equation}
where the coefficients $A,B,C,D$ are functions only of $q$. In detail the constraint \eqref{fgrelation} requires
$$ (AD)_q=2(e^q-e^{-q}),\quad (BC)_q=2(e^q-e^{-q}), \quad A_q C=AC_q,\quad B_qD=BD_q,$$
and hence $$C=\alpha A, \quad D=\alpha  B, \quad \alpha AB=2(e^q+e^{-q})+2\gamma,$$
where $\alpha$ and $\gamma$ are constants.
Since $\lambda$ can be shifted by a function of $q$ without causing an essential change, there is a family of equivalent solutions to these constraints and it is a matter of convenience which choice is most suitable. For future purposes, it also turns out to be useful to define
$$\gamma = (e^{2\tau}+e^{-2\tau}).$$
A representative choice for $f$ and $g$ that will be used below is
\begin{eqnarray}\label{sgfandg}
  \nn
\nn  f &=& \frac{1}{\sigma}\left(2e^{p-\lambda} + e^{-p+\lambda}\left( e^q+e^{-q}+\gamma\right)\right),\\
g &=& \sigma\left( e^{\lambda}\left(e^q+e^{-q}+\gamma\right)+2e^{-\lambda}\right).
\end{eqnarray}
Using these, the defect conditions \eqref{defectcondition} can be rewritten in terms of $p, q$ and $\lambda$ as follows:
\begin{eqnarray}\label{sgpqconditions}
p_x-p_t+2\lambda_t&=&-\frac{\sigma}{2}\, e^{\lambda}(e^{q}-e^{-q})-\frac{1}{2\sigma}e^{-p+\lambda}
(e^{q}-e^{-q}),\nn\\
q_x-q_t&=&-\frac{\sigma}{2}\,
\left(e^{\lambda}
(e^{q}+e^{-q}+\gamma)-2e^{-\lambda}\right),\nn\\
q_x+q_t&=&\frac{1}{2\sigma}\,\left(e^{-p+\lambda}
(e^{q}+e^{-q}+\gamma)-2e^{p-\lambda}\right).
\end{eqnarray}
For the sinh-Gordon model, the static solution in the bulk is $u=v=0$ and this satisfies the defect conditions \eqref{sgpqconditions} provided
\begin{equation}
\nn e^{2\lambda}=\frac{1}{2\cosh^2\tau}.
 \end{equation}
 On the other hand, purely imaginary solutions to the sinh-Gordon model are the solutions to the sine-Gordon model, the least energy static solutions in the bulk correspond to $u=2\pi i a$ and $v=2\pi i b$ where $a$ and $b$ are integers, and the defect conditions permit $a\ne b$ provided $\lambda$ is chosen suitably. In fact, the conditions imply:
\begin{equation}\label{staticlambda}
e^{2\lambda}=\left\{\begin{array}{cc}
                        1/2\cosh^2\tau & \hbox{if}\ a-b\ \hbox{is\ even} \\
                       1/2\sinh^2\tau  & \hbox{if}\ a-b\ \hbox{is\ odd}.
                      \end{array}\right.
\end{equation}

\subsection{The Liouville equation}

The Liouville field theory fits into the same scheme by truncating the previous choices for $f$ and $g$ in the sinh/sine-Gordon model found in \eqref{sgfandg}. Thus, for example,
\begin{eqnarray}\label{Lpotential}
  \nn
\nn  f &=& 2e^{p-\lambda}\\
g &=& e^{\lambda}\left(e^q+e^{-q}+\gamma\right),
\end{eqnarray}
is an adequate choice since
$$\frac{1}{2}(f_q g_\lambda - f_\lambda g_q)=e^{p+q}-e^{p-q}.$$
In this case, there is no place for an arbitrary parameter to correspond to $\sigma$ since any such could be removed by a translation of $\lambda$. On the other hand, the parameter $\gamma$ can be chosen freely.

\p Further, dropping one or other of the exponential pieces $e^q$ (or $e^{-q}$) in $g$ leads to a defect that couples the Liouville model for $u$ (or $v$) to free massless field theory for $v$ (or $u$).

\subsection{The Tzitz\'eica equation}

\p For the Tzitz\'eica model the bulk potentials are,
$$U=e^{2p+2q}+2e^{-p-q}=e^{2u}+2e^{-u}, \quad V=e^{2p-2q}+2e^{-p+q}=e^{2v}+2e^{-v},$$
and the most general ansatz is
\begin{equation}
f=A e^{2p-2\lambda}+B e^{-p+\lambda}, \quad g=C e^{2\lambda}+ D e^{-\lambda},
\end{equation}
with the coefficients $A,B,C,D$  being functions only of $q$.
The constraints  following from \eqref{fgrelation} are
$$A_qD=2AD_q,\quad \quad 2B_qC=BC_q ,\quad (AC)_q=(e^{2q}-e^{-2q}),\quad (BD)_q=4(e^q-e^{-q}),$$
for which the general solution is
$$BD=4(e^q-e^{-q}),\quad A=\alpha D^2, \quad C=\frac{B^2}{32\alpha}.$$
It is always possible to shift $\lambda$ by a function of $q$ and, for example, $A$ (and therefore $D$) can be chosen to be constants. Using a further shift one of these constants may be removed and a convenient expression for the most general solution up to these translations of $\lambda$ is:
\begin{eqnarray}\label{Tpotential}
  \nn
\nn  f &=& \frac{1}{\sigma}\left(e^{2p-2\lambda} + e^{-p+\lambda}\left( e^q+e^{-q}\right)\right),\\
g &=& \frac{\sigma}{2}\left(8e^{-\lambda} +  e^{2\lambda}\left(e^q+e^{-q}\right)^2\right).
\end{eqnarray}
This contains one free parameter $\sigma$.

\subsection{Massive free fields}

It is also instructive to consider the case where the fields to either side of the defect are free (and massive with mass parameter $m$). In this situation, similar considerations lead to
\begin{eqnarray}\label{KGpotential}
  \nn
\nn  f &=& m\left(\frac{(p-\lambda)^2}{\beta} +\alpha q^2\right),\\
g &=&m\left(\frac{\lambda^2}{\alpha} +\beta q^2\right) ,
\end{eqnarray}
where $\alpha$ and $\beta$ are undetermined parameters.

\p
One question is whether both of these parameters are effective after $\lambda$ is eliminated (or, equivalently, integrated out in a functional integral). After some algebra, the result for the defect part of the Lagrangian (after removing a total time derivative) is the following:
\begin{equation}\label{integratedpotential}
{\cal L}_D=\delta(x)\left[\frac{4\alpha\beta}{m(\alpha+\beta)}\, q_t^2-\half\left(\frac{\alpha- \beta}{\alpha+\beta}\right)(uv_t-vu_t)-m\left(
\frac{p^2}{\alpha+\beta} +(\alpha+\beta) q^2\right)\right].
\end{equation}
This still depends upon two parameters, yet in an interesting manner. For example, the limit $\alpha\rightarrow 0$ gives the free field type I defect considered in an earlier article \cite{bczlandau}, as does the limit $\beta\rightarrow 0$, apart from an inessential sign change in the term linear in time derivatives. From this observation it is clear that the new framework does indeed engender an alternative type of defect to those considered previously. However, it is not straightforward to eliminate $\lambda$ in the other, nonlinear, examples.

\p The expressions for $f$ and $g$ in the sinh/sine-Gordon model given in \eqref{sgfandg} also contain two free parameters and it is to be expected these survive in the quadratic limit regarded as an expansion about a classical constant configuration. One way to facilitate the limit  is to put $\sigma=e^{\eta}$, and note an alternative but quite symmetrical expression for $\D$:
$$\D=4\sqrt{2}\left(e^{-\lambda+p/2}\cosh\frac{p-2\eta}{2}\cosh\frac{q+2\tau}{2}
+e^{\lambda-p/2}\cosh\frac{p+2\eta}{2}\cosh\frac{q-2\tau}{2}\right),$$
which may be expanded about the point $p=q=\lambda=0$. After shifting $$\lambda\rightarrow \lambda+\frac{q\tanh\tau}{2},$$
the quadratic form is diagonal and resembles \eqref{KGpotential}; putting $m=\sqrt{2}$,  $\alpha$ and $\beta$  are given by
$$\alpha=\frac{\sigma}{2\cosh\tau} ,\quad \beta=\frac{1}{2\sigma\cosh\tau}.$$
These parameters lie on the set of curves
$$\alpha\beta=\frac{1}{4\cosh^2\tau}.$$
On the other hand, the
quadratic limit of the expression \eqref{Tpotential} giving the functions $f$ and $g$ for the Tzitz\'eica equation is a  particular one parameter set within the general two parameter family. Thus, for the Tzitz\'eica equation ($m=\sqrt{6}$) one finds:
\begin{equation}
\alpha=\frac{1}{\sqrt{6}\sigma},\quad \beta = \frac{2\sigma}{\sqrt{6}},
\end{equation}
corresponding to points on the curve $ \alpha\beta = \frac{1}{3}$.

\p If a plane travelling wave,
$$u=e^{-i\omega t}(e^{ikx}+Re^{-ikx}),\quad v= e^{-i\omega t} \, Te^{ikx},\quad \omega =m\cosh\theta,\ k=m\sinh\theta,$$
encounters a defect with the potential \eqref{integratedpotential} then there is no reflection ($R=0$), and the transmission factor $T$ is given by:

\begin{equation}\label{freetransmission}
T=\frac{i\left(\alpha e^\theta - \beta e^{-\theta}\right)+1}{i\left(\alpha e^\theta - \beta e^{-\theta}\right)-1}\, .
\end{equation}
One difference from the previously considered cases (with $\alpha=0$ or $\beta=0$) is the possibility of a `bound state' when $\alpha=\beta$, for example of the form $$u=u_0\cos\omega t\, e^{m\zeta x},\ x<0;\quad v=0,\ x>0,\ \zeta=-1/2\alpha,$$
with the constraint $\alpha<-1/2$.
The contributions to the energy of this solution from the bulk and defect exactly cancel, though both are time-dependent, leading to a zero
energy excitation degenerate with the constant `vacuum' (in which all fields are zero everywhere).

\p Since the present scheme can accommodate all the known single field integrable Toda systems one might be optimistic that a generalisation of the scheme will encompass all Toda models, conformal or affine, irrespective of the choice of root data. At this time, however, this generalisation, if it exists, is not known.

\section{A single soliton passing a defect}

\p So far, nothing has been said about integrability. Nevertheless, this new class of defect is thought to be integrable on the basis of some indirect evidence. For example, if this is the case, at the very least single solitons for both the sine-Gordon model and the complex Tzitz\'eica model are expected to pass safely through a defect suffering at most a delay. In this section, the behaviour of  single soliton solutions for these two models will be explored. In addition, in appendix \ref{appendixA} an energy-like spin 3 charge for the sine-Gordon model is calculated and found to be conserved on using the defect conditions \eqref{sgpqconditions}. Ideally, a Lax pair formulation is needed to generalise the ideas presented in \cite{bczlandau}.

\subsection{The sine-Gordon soliton}

\p In the previous section the sinh/sine-Gordon model were considered together but solitons are real solutions of sine-Gordon or purely imaginary solutions of the sinh-Gordon equation. For ease of notation, and compatibility with earlier sections, the fields $u$ and $v$ will be pure imaginary.
Then the defect conditions \eqref{sgpqconditions} will determine how a soliton scatters with the defect.
The defect parameters will be taken to be real.

\p In a situation where the intial defect has either no discontinuity, or a discontinuity proportional to $4\pi$, a single soliton solution can be written as follows:
$$
e^{u/2}=\frac{1+ E}{1- E},\quad E=e^{ ax+bt+c}, \quad a=\sqrt{2}\cosh\theta,\quad b=-\sqrt{2}\sinh\theta,
\quad e^{v/2}=\frac{1+z E}{1-z E},
$$
where $z$ represents the delay, the rapidity $\theta>0$ indicates a soliton travelling from left to right along the $x$-axis, and $e^c$ is purely imaginary. Replacing $E\rightarrow -E$ (or equivalently shifting $c\rightarrow c+i\pi$) provides an expression for an anti-soliton.

\p
The final pair of defect conditions  \eqref{sgpqconditions} do not involve $\lambda_t$ and can be used to  obtain two expressions for $\lambda$,
\begin{eqnarray}
e^{\lambda}&=&-\frac{2}{\sigma}\,\,\frac{\sigma^2(q_x+q_t)+e^p(q_x-q_t)}{(e^p-e^{-p})
(e^{q}+e^{-q}+\gamma)}
\label{exp1}\\ \nn &&\\
e^{-\lambda}&=&-\frac{1}{\sigma}\,\,\frac{\sigma^2(q_x+q_t)+e^{-p}(q_x-q_t)}{(e^p-e^{-p})
}\label{exp2}.
\end{eqnarray}
These two expressions must be consistent and  will determine both $z$ and $\lambda$. In fact, there will be two possible choices for $z$ corresponding to the $i\pi$ ambiguity in the possible static solutions for $\lambda$ given by \eqref{staticlambda}. Explicitly, the two possibilities for the delay are given by $z=z_1$ or $z=z_2$, where
\begin{equation}\label{sgdelay}
z_1=\tanh\left(\frac{\theta-\eta+\tau}{2}\right)\,\tanh\left(\frac{\theta-\eta-\tau}{2}\right),
\quad z_2=1/z_1, \quad \sigma = e^\eta.
\end{equation}
For $z=z_1$, the companion expression for $\lambda$ is given by
\begin{equation}\label{sglambda}
e^{\lambda_1}=\frac{1}{\sqrt{2}\cosh\tau}\,\, \frac{(1+E_0)(1+zE_0)}{(1+\rho E_0)(1+\tilde\rho E_0)},\quad \rho=\tanh\left(\frac{\theta-\eta+\tau}{2}\right), \ \tilde\rho=\tanh\left(\frac{\theta-\eta-\tau}{2}\right),
\end{equation}
where $E_0 =E(0,t)$, and there is a similar expression for $\lambda_2$.

\p
Interestingly, the expression \eqref{sgdelay} indicates that the delay is identical to a delay that would be experienced by a soliton passing through two defects of type I (see for example \cite{bczsg05}) with parameters $\eta\pm\tau $. Because $E_0$ is purely imaginary the expression \eqref{sglambda} for $\lambda_1$  indicates that $\lambda_1$ is complex and nowhere singular as a function of real $t$. In order to decide which of the two solutions should be chosen the starting value of $\lambda$ (that is, the value $\lambda$ has when the soliton is far away but approaching the defect) needs to be specified - effectively, the defect has two states associated with it even when the static field configurations to either side of it are $u=0$ and $v=0$. The modulus of $e^{\lambda_1}$, with $0<|z_1|\le 1$, grows  to a maximum at $E_0^4=z_1^{-2}$ then falls to its initial value. On the other hand, the phase of $e^{\lambda_1}$ is more interesting since it is the product of four terms, each having a single soliton (or anti-soliton) form (but is a function only of time):
\begin{equation}
e^{2i{\rm Im}\lambda_1}=\left(\frac{1+E_0}{1-E_0}\right)\left(\frac{1+z_1 E_0}{1-z_1 E_0}\right)
\left(\frac{1-\rho E_0}{1+\rho E_0}\right)\left(\frac{1-\tilde\rho E_0}{1+\tilde\rho E_0}\right).
\end{equation}
The first factor (provided $E_0$, which is pure imaginary, has a positive imaginary part) has a phase whose angle monotonically decreases by $\pi$ as $t$ runs over its range $(-\infty,\infty)$. On the other hand, if the imaginary part of $E_0$ had been negative, the phase angle would have increased by $\pi$. So, the total effect of the four terms will be either zero (if not more than one of $\rho$ or $\tilde\rho$ is negative), or $-4\pi$ (if both $\rho$ and $\tilde\rho$ are negative). Thus the phase angle of $e^{\lambda_1}$ either shifts by $0$ or $-2\pi$. The case where the imaginary part of $\lambda_1$ shifts by $-2\pi$ is quite interesting. There, the ingoing soliton emerges as a soliton but only after flipping to an anti-soliton and back again, in a virtual sense, since that is what would have happened had the soliton passed two separated defects with the chosen parameters. In other words, keeping track of $\lambda$ distinguishes the two possible cases ($z_1>0$) where a soliton emerges as a soliton. In the other two cases (one of $\rho$ or $\tilde\rho$ is negative), the soliton emerges as an antisoliton.

\p As was the case with type I defects, and as indicated above, the delay $z$ can indicate a change in the character of the soliton as it passes (if $\eta - \tau<\theta<\eta+\tau$, then $z_1<0$, and an approaching soliton will emerge as an anti-soliton), or the soliton may be absorbed (if $\theta =\eta-\tau$ or $\theta=\eta +\tau$, meaning $z_1=0$). In the latter case, the expression for $\lambda$ interpolates `even' and `odd' static solutions given by \eqref{staticlambda}, as it should since the defect stores the topological charge (and the energy-momentum and other charges) transported by the soliton. The limit $\tau\rightarrow 0$ is interesting because in that limit the defect (at least as far as the scattering property is concerned) is behaving like another soliton of rapidity $\eta$. This lends a little more credibility to the idea (already mentioned in \cite{bczsg05}) that a pair of defects with the same parameter behaves like a soliton. These results are very suggestive of the idea that at least for the sine-Gordon model the type II defects are `squeezed', or `fused', pairs of type I defects.

\p Finally, it is not difficult to check directly that the first of the three defect conditions \eqref{sgpqconditions} is satisfied by the soliton solution without any further constraints on $\lambda$ or $z$.

\p A question that will not be addressed here is how the type II defect should be described by a transmission matrix in the quantum sine-Gordon field theory. Presumably, a generalisation of the Konik-LeClair transmission matrix (see \cite{bczsg05}) will need to be found and this will be postponed for a future investigation.

\subsection{The Tzitz\'eica equation}

\p The solitons for the Tzitz\'eica equation can be analysed similarly although in this case the soliton  is complex (though its energy and momentum are real). Using the same conventions as before with the potential associated with the choice \eqref{Tpotential},
the defect conditions are:
\begin{eqnarray}\label{Tpqconditions}
p_x-p_t+2\lambda_t&=&-\frac{\sigma}{2}\, e^{2\lambda}(e^{2q}-e^{-2q})-\frac{1}{2\sigma}\, e^{-p+\lambda}(e^{q}-e^{-q}),\nn\\
q_x-q_t&=&-\frac{\sigma}{2}\left(e^{2\lambda}(e^{q}+e^{-q})^2-4e^{-\lambda}\right),\nn\\
q_x+q_t&=&\frac{1}{2\sigma}\left(e^{-p+\lambda}(e^{q}+e^{-q})-2e^{2p-2\lambda}\right).
\end{eqnarray}
Single soliton solutions in the bulk are given by the expressions \cite{Mikhailov81,Cherdantzev90,Mackay93}
$$
e^{u}=\frac{(1+  E)^2}{(1-4  E+E^2)},\quad
 e^{v}=\frac{(1+z E)^2}{(1-4 z E+z^2 E^2)},
$$
with
$$ E=e^{ ax+bt+c}, \quad a=\sqrt{6}\cosh\theta,\quad b=-\sqrt{6}\sinh\theta,$$
where $z$ represents the delay of the outgoing soliton. The constant $e^c$ is chosen so that the expressions for $u$ and $v$ are nonsingular for all real choices of $t$ and $x$. The last two of the defect conditions \eqref{Tpqconditions} can be regarded as a pair of cubic equations for $\Lambda\equiv e^\lambda$ of the
form
\begin{equation}
 \alpha_1\Lambda^3 +\beta_1\Lambda^2+\gamma_1=0,\quad \alpha_2\Lambda^3+\beta_2\Lambda+\gamma_2=0,
\end{equation}
where the coefficients depend upon $p,\ \sigma$,\ $q$ and the derivatives of $q$.
Together, these may be solved to give
\begin{equation}
 \Lambda=\frac{\alpha_1\beta_2^2\gamma_1+\beta_1\gamma_2(\alpha_1\gamma_2-\alpha_2\gamma_1)}{\alpha_2\beta_1\beta_2\gamma_1-(\alpha_1\gamma_2-\alpha_2\gamma_1)^2},\quad \frac{1}{\Lambda}=\frac{\alpha_2\beta_1^2\gamma_2+\alpha_1\beta_2(\alpha_1\gamma_2-\alpha_2\gamma_1)}{\alpha_2\beta_1\beta_2\gamma_1-(\alpha_1\gamma_2-\alpha_2\gamma_1)^2}.
\end{equation}
Demanding these two expressions are compatible and inserting the soliton solutions reveals, after some algebra, three possibilities for $z$:
\begin{eqnarray}
z_1&=& \frac{(e^{-\theta+\eta}+e^{i\pi/6})(e^{-\theta+\eta}+e^{-i\pi/6})}
{(e^{-\theta+\eta}-e^{i\pi/6})(e^{-\theta+\eta}-e^{-i\pi/6})},\quad e^{\eta}=\sqrt{2}\sigma\nn\\
z_2=\bar z_3&=& \frac{(e^{-\theta+\eta}-e^{i\pi/6})(e^{-\theta+\eta}+e^{-i\pi/2})}
{(e^{-\theta+\eta}+e^{i\pi/6})(e^{-\theta+\eta}-e^{-i\pi/2})}.
\end{eqnarray}
These may also be rewritten more suggestively:
\begin{eqnarray}
z_1&=&\coth\left(\frac{\theta-\eta}{2}-\frac{i\pi}{12}\right)
\coth\left(\frac{\theta-\eta}{2}+\frac{i\pi}{12}\right),\\
z_2=\bar z_3&=&\coth\left(\frac{\theta-\eta}{2}+\frac{i\pi}{4}\right)
\tanh\left(\frac{\theta-\eta}{2}-\frac{i\pi}{12}\right),
\end{eqnarray}
and
$$z_1z_2z_3=1.$$
Finally, as examples, for the two cases $z=z_1$ or $z=z_2$ expressions for the field $\lambda$ are
\begin{eqnarray}
e^{\lambda_1}&=&\frac{(1+E_0)(1+z_1\, E_0)}{(1+2\,\rho_1\,E_0+z_1 \,E_0^2)},\phantom{\,e^{-2i\pi/3}}
\quad \rho_1=\frac{(e^{-\theta+\eta}-\sqrt{2})(e^{-\theta+\eta}+\sqrt{2})}
{(e^{-\theta+\eta}-e^{i\pi/6})(e^{-\theta+\eta}-e^{-i\pi/6})},\label{lambdaT1}\\
e^{\lambda_2}&=&\frac{(1+E_0)(1+z_2\, E_0)}{(1+2\,\rho_2\, E_0+z_2 \,E_0^2)}\,e^{-2i\pi/3},
\quad \rho_2=\frac{(e^{-\theta+\eta}-\sqrt{2}e^{i\pi/3})(e^{-\theta+\eta}+\sqrt{2}e^{i\pi/3})}
{(e^{-\theta+\eta}+e^{i\pi/6})(e^{-\theta+\eta}-e^{-i\pi/2})}\label{lambdaT2}.
\end{eqnarray}
For $z=z_3$ the corresponding formulae are the complex conjugates of the expressions in \eqref{lambdaT2}.
 The possible asymptotic values of $u$ and $v$ for  soliton solutions are $u=2\pi ia$, $v=2\pi ib$, and the corresponding asymptotic values of $\lambda$ required by the defect conditions are $\lambda=2\pi ic$ or $\lambda=\pm 2\pi i/3+2\pi ic$ with $a,b,c$ integers. The formulae \eqref{lambdaT1} and \eqref{lambdaT2} for $\lambda$ provide examples of this. Once again, as was found to be the case for the sine-Gordon model, part of the specification of the defect must be the initial choice of $\lambda$ (essentially, for the soliton, one of three).

\section{Defects and B\"{a}cklund transformations}

\p In \cite{bczlandau} it was pointed out using several examples that integrable defect conditions for type I defects coincide with B\"{a}cklund transformations `frozen' at the defect location. This impression was strongly reinforced by subsequent analysis of the $a_n^{(1)}$ affine Toda models \cite{bcztoda, cz07}. However, it was also found that while the $a_2^{(2)}$ Toda model has B\"acklund transformations these cannot be used directly to construct integrable defects within the type I scheme.  At first sight this seemed puzzling and the purpose of this section is to show how a `folding' procedure \cite{mikhailov79, OT83fold} may be used to obtain a B\"{a}cklund transformation for the Tzitz\'eica model, making use of  two similar, yet different, sets of defect conditions obtained in \cite{bcztoda} for the $a_2^{(1)}$ Toda model.

\p First a little background is necessary. The equation of motion for an $a_2^{(1)}$ Toda field $u$ is
\begin{equation}\label{a21em}
\partial^2 u=-2\sum_{j=0}^2\,\alpha_j\,e^{\alpha_j\cdot u},
\end{equation}
 where, with respect to a basis of orthonormal vectors $\{e_0, e_1, e_2\}$ in a three dimensional  Euclidean space, the $a_2^{(1)}$ roots are:
\begin{equation}\label{newnotation}
\alpha_1=e_1-e_2,\quad \alpha_2=e_2-e_0, \quad \alpha_0=e_0-e_1.
\end{equation}
The projections of the field $u$ onto the orthonormal basis are $u_0, u_1, u_2$ satisfying the constraint $u_0+u_1+u_2=0$.
From \eqref{a21em} it follows that the corresponding equations for the projections read
\begin{equation}\label{a21emp}
\partial^2u_j=-2(e^{u_j-u_{j+1}}-e^{- u_j+u_{j-1}}),\quad j=0,1,2,
\end{equation}
where the subscripts on the right hand side are to be understood modulo 3.
Then, the folding procedure consists of setting one of the fields to zero, for instance $u_2=0$ (i.e. $u_1=-u_0$), to obtain the Tzitz\'eica equation of motion with the same normalisations as had been assumed when writing down the Tzitz\'eica potential in section 4.2. Note, the alternative choices $u_0=0$ or $u_1=0$ would lead to the same conclusion.
The defect conditions that must hold at the defect ($x=x_0$) when sewing together two $a_2^{(1)}$ Toda fields $u$ and $\lambda$ are
\begin{equation}\label{a21dc}
\partial_{x}u-A \pt u-B \pt \lambda+ {\cal D}_u=0, \quad \partial_{x}\lambda-B^{T}\pt u +A\pt \lambda-{\cal D}_\lambda=0, \quad B=(1-A),
\end{equation}
with
\begin{equation}\label{a21dp}
{\cal D}=\sqrt{2}\,\sum_{j=0}^2\left(\sigma\,
e^{\alpha_j\cdot(B^T u+B \lambda)/2}+\frac{1}{\sigma}\, e^{\alpha_j \cdot B(u-\lambda)/2}\right),
\quad B=2\sum_{a=0}^2 w_a\left( w_a-w_{a+1}\right)^T,
\end{equation}
where $w_1, w_2$ ($w_3\equiv w_0=0$) are the fundamental highest weights of the Lie algebra $a_2^{(1)}$.
By using similar notation as in \eqref{a21emp} for the two fields $u$ and $\lambda$ and light-cone coordinates $x_{\pm}=(t\pm x)/2$, the full set of defect conditions \eqref{a21dc} read
\begin{eqnarray}\label{dctypeI}
\partial_+(u_1-u_2)-\partial_+(\lambda_1-\lambda_2)&=&\sqrt{2}\,\sigma(e^{u_0-\lambda_1}-2e^{u_1-\lambda_2}
+e^{u_2-\lambda_0}),\nn\\
\partial_+(u_2-u_0)-\partial_+(\lambda_2-\lambda_0)&=&\sqrt{2}\,\sigma(e^{u_1-\lambda_2}-2e^{u_2-\lambda_0}+e^{u_0-\lambda_1}),\nn\\
\partial_+(u_0-u_1)-\partial_+(\lambda_0-\lambda_1)&=&\sqrt{2}\,\sigma(e^{u_2-\lambda_0}-2e^{u_0-\lambda_1}+e^{u_1-\lambda_2}),\nn\\
&&\nonumber\\
\partial_-(u_1-u_2)-\partial_-(\lambda_2-\lambda_0)&=&\sqrt{2}\,\sigma^{-1}(2e^{-u_2+\lambda_2}-e^{-u_1+\lambda_1}-e^{-u_0+\lambda_0}),\nn\\
\partial_-(u_2-u_0)-\partial_-(\lambda_0-\lambda_1)&=&\sqrt{2}\,\sigma^{-1}(2e^{-u_0+\lambda_0}-e^{-u_2+\lambda_2}-e^{-u_1+\lambda_1}),\nn\\
\partial_-(u_0-u_1)-\partial_-(\lambda_1-\lambda_2)&=&\sqrt{2}\,\sigma^{-1}(2e^{-u_1+\lambda_1}-e^{-u_0+\lambda_0}-e^{-u_2+\lambda_2}).
\end{eqnarray}
In the bulk, the expression \eqref{dctypeI} would be the B\"acklund transformation discovered by Fordy and Gibbons \cite{Fordy80}.

\p Unfortunately, the folding procedure cannot be applied directly to the defect conditions because they are simply incompatible with folding. This fact can be expressed heuristically by noting that  the defect conditions \eqref{dctypeI} do not treat solitons and antisolitons identically (a feature already pointed out in \cite{bcztoda, cz07} and expected since solitons and anti-solitons are associated with different representations of the $a_2$ algebra ), because each type of soliton experiences a different delay on transmission through the defect.  The soliton solution of the Tzitz\'eica model can be thought of as a particular soliton-antisoliton solution of the $a_2^{(1)}$ affine Toda
model, and, since the components of a multi-soliton are delayed independently by the defect, its components will be treated differently by \eqref{dctypeI}. Therefore, the Tzitz\'eica soliton cannot survive intact. A remedy is provided by observing  that an alternative defect setting is available if the matrix $B$ in \eqref{a21dp} is replaced by its transpose. The resulting set of defect conditions describes a system, which is still integrable yet interchanges the delays experienced by a soliton or antisoliton when compared with the previous case.  This suggests that two different types of defect, one built using $B$ (at $x=x_0$) and the other with $B^T$ (at $x=x_1$), then `squeezed' together ($x_1\rightarrow x_0$), might allow the folding procedure to be applied successfully. The second set of defect conditions matches two $a_2^{(1)}$ fields $\lambda$ and $v$ and would be written in a similar manner to \eqref{dctypeI} but using $B^T$ instead of $B$.

\p Since the incoming ($u$) and outgoing ($v$) solitons are required to satisfy the Tzitz\'eica equation of motion, the projections $u_2, v_2$ can be set equal to zero.  Consequently, the field $\lambda$ is forced to satisfy the following constraint (at $x=x_0$):
\begin{equation}\label{constraint}
2\,e^{-\lambda_0}=e^{-p_0+\lambda_2}\,(e^{q_0}+e^{-q_0}), \quad p_0=\frac{u_0+v_0}{2}, \quad q_0=\frac{u_0-v_0}{2}.
\end{equation}
Setting $u_0\equiv u,\ v_0\equiv v,\ \lambda_2\equiv -\lambda$  and sending $\sigma\rightarrow1/(\sqrt{2}\,\sigma)$ the two sets of defect conditions lead to
\begin{eqnarray}\label{BTTT}
\partial_-(p-\lambda)&=&\frac{\sigma}{2}\,e^{2\lambda}(e^{2q}-e^{- 2q})\label{BTT1}\\
\partial_+\lambda&=&-\frac{1}{2\sigma} e^{-p+\lambda}(e^{ q}-e^{- q}),\label{BTT2}\\
\partial_-q&=&\frac{\sigma}{2}\,(e^{2\lambda}(e^{ q}+e^{- q})^2-4 e^{-\lambda}),\label{BTT3}\\
\partial_+q&=&
\frac{1}{2\sigma}(e^{-p+\lambda}(e^{ q}+e^{- q})-2 e^{2p-2\lambda}).\label{BTT4}
\end{eqnarray}
 If, instead of being `frozen' at $x=x_0$, equations \eqref{BTTT}-\eqref{BTT4} were required to hold in the bulk, they do, in fact, represent a B\"{a}cklund transformation for the Tzitz\'eica equation. This can be seen by cross-differentiating the expressions \eqref{BTT3} and \eqref{BTT4} to eliminate $\lambda$, to find that if the field $u$ satisfies the Tzitz\'eica equation then the field $v$ also satisfies it. Also, cross-differentiating expressions \eqref{BTT1} and \eqref{BTT2} an equation of motion satisfied by the field $\lambda$ emerges:
\begin{equation}\label{a22em}
\partial^2\lambda=-(e^{q}+e^{-q})\,e^{\lambda-p}(e^{2\lambda}-e^{-\lambda}).
\end{equation}
Inevitably, this depends on the fields $u$ and $v$. The B\"acklund transformation \eqref{BTTT}-\eqref{BTT4} seems not to have been reported elsewhere in the literature \cite{TzitzeicaB1,TzitzeicaB2,TzitzeicaB3}.

\p On the other hand, since equations (\ref{BTTT}-\ref{BTT4}) are supposed to hold only at $x=x_0$, and because the quantity $\lambda$ is  confined at $x=x_0$ and depends only on $t$, the sum of the pair \eqref{BTT1} and \eqref{BTT2}, together with \eqref{BTT3} and \eqref{BTT4} are precisely the three defect conditions \eqref{Tpqconditions}. Hence, for the type II defect, the number of defect conditions following from the Lagrangian \eqref{lambdalagrangian} is one less than the number of equations specifying the B\"acklund transformation described above. This is quite different to the previous situation where the Lagrangian description of a type I defect led directly to the  frozen B\"acklund transformation (and hence to the B\"acklund transformation itself).

\p Clearly, using the same idea, the defect conditions \eqref{sgpqconditions} can be augmented to obtain an alternate B\"{a}cklund transformation for the sine-Gordon model that depends on two parameters:
\begin{eqnarray}
\partial_-(p-\lambda)&=&\frac{\sigma}{2}\,e^{\lambda}(e^{q}-e^{- q})\nn\\
\partial_+\lambda&=&-\frac{1}{2\sigma} e^{-p+\lambda}(e^{ q}-e^{- q}),\nn\\
\partial_-q&=&\frac{\sigma}{2}\,(e^{\lambda}(e^{ q}+e^{- q}+\gamma)-2 e^{-\lambda}),\nn\\
\partial_+q&=&
\frac{1}{2\sigma}(e^{-p+\lambda}(e^{ q}+e^{- q}+\gamma)-2 e^{(p-\lambda)}).
\end{eqnarray}
From these relations, in a similar manner as previously, the equations of motion for the sine-Gordon fields $u$ and $v$ are recovered and the field $\lambda$ satisfies,
\begin{equation}
\partial^2\lambda=-\frac{1}{4}\,e^{-p}\,(4\,e^{2\lambda}-
(e^{q}+e^{-q})(2-\gamma\,e^{2\lambda})).
\end{equation}

\p The fact that there appear to be generalisations of the defect conditions, which are only indirectly related to B\"acklund transformations, and yet likely to be integrable, generates a sense of optimism that the framework will generalise to encompass all affine Toda models.

\section{Defects as Hamiltonian constraints}

\p So far, properties of defects, and the relationship of the defect conditions to the conservation of a suitably defined momentum, have been derived from first principles from a Lagrangian starting point. It is interesting to ask if the framework can be formulated within a Hamiltonian setting. In this section this will be attempted, at least at a formal level, by explaining the main ideas, albeit sketchily. The setup demonstrates explicitly that the presence
of a defect reduces the independent degrees of freedom of the system in the sense of providing defect conditions that can be regarded as a set of constraints on the fields
$u$ and $v$ (for type I defects), or $u$, $v$ and $\lambda$ (for type II defects).
This fact is highlighted by the emergence of second class constraints
in the Hamiltonian (for a detailed description of these, see for example \cite{HenneauxConstraints}).

\p The
discussion can begin by considering a system with a type I defect.  In this case, the starting point is the following Lagrangian
density
\begin{equation}
\mathcal{L}=\theta(-x)\mathcal{L}_u+\theta(x)\mathcal{L}_v+\delta(x)\left(\frac{u\,
v_t-v\, u_t}{2}-\mathcal{D}(u, v)\right),
\end{equation}
with
\begin{equation}
\mathcal{L}_u=\frac{1}{2}\partial_\mu u\,\partial^\mu u-U(u),\quad
\mathcal{L}_v=\frac{1}{2}\partial_\mu v\,\partial^\mu v-V(v).
\end{equation}
According to the usual definitions, and treating the theta and delta functions formally, the canonical momenta conjugate to the fields $u$ and $v$ are,
\begin{eqnarray}\label{momenta}
\pi_u&=&\frac{\partial \mathcal{L}}{\partial
u_t}=\theta(-x)\,u_t-\delta(x)\,\frac{v}{2},\nn\\
\pi_v&=&\frac{\partial \mathcal{L}}{\partial
v_t}=\theta(x)\,v_t+\delta(x)\,\frac{u}{2}.
\end{eqnarray}
 By comparison with what happens within each half line, $x<0$ or $x>0$, the canonical momenta
are not well-defined at the defect location. In other words, at $x=0$
it is not possible to write the time derivatives of the fields
(Lagrangian variables) in terms of the canonical momenta (Hamiltonian
variables). At $x=0$ the canonical momenta are not independent, and
the definitions \eqref{momenta} provide constraints amongst the canonical variables. These are
\begin{equation}\label{Hconstraints}
\chi_1=\pi_u+\frac{v}{2}=0,\quad \chi_2=\pi_v-\frac{u}{2}=0;
\end{equation}
these are primary constraints.
The Hamiltonian is given by
\begin{equation}\label{H}
H=\int_{-\infty}^\infty dx\,\mathcal{H}
\end{equation}
with
\begin{equation}\label{Hd}
\mathcal{H}=\theta(-x)\left(\frac{\pi_u^2+u_x^2}{2}+U\right)+
\theta(x)\left(\frac{\pi_v^2+v_x^2}{2}+V\right)+\delta(x)\left(\mathcal{D}+\mu_1\chi_1+\mu_2\chi_2\right),
\end{equation}
where $\mu_1$ and $\mu_2$ are functions of the fields $u,v$ together
with their momenta.
They can be determined by using the fact that the constraints $\chi_1$
and $\chi_2$ must be preserved in time. In other words, the relations
\begin{equation}\label{constraintsintime}
{\chi_1}\,_t=\{\chi_1,H\}=0,\quad
{\chi_2}\,_t=\{\chi_2,H\}=0,
\end{equation}
must hold. The Poisson bracket of two
functionals $F=\int_{-\infty}^{\infty}dx\, \mathcal{F}$ and
$G=\int_{-\infty}^{\infty}dx\, \mathcal{G}$ is defined formally as follows
\begin{equation}\label{PB}
\{F,G\}=\int_{-\infty}^{\infty}\,dx\left(\frac{\delta F}{\delta
u}\frac{\delta G}{\delta \pi_u}-\frac{\delta F}{\delta
\pi_u}\frac{\delta G}{\delta u}\right)+\int_{-\infty}^\infty
\,dx\left(\frac{\delta F}{\delta v}\frac{\delta G}{\delta
\pi_v}-\frac{\delta F}{\delta \pi_v}\frac{\delta G}{\delta v}\right).
\end{equation}
Using this definition and the Hamiltonian \eqref{H}, for which,
\begin{eqnarray}\label{HCanEqu}
\frac{\delta H}{\delta \pi_u}&\equiv& u_t=\frac{\partial\mathcal{
H}}{\partial \pi_u}=\theta(-x)\pi_u+\delta(x)\mu_1,\ \
\frac{\delta H}{\delta \pi_v}\equiv v_t=\frac{\partial\mathcal{
H}}{\partial \pi_v}=\theta(x)\pi_v+\delta(x)\mu_2,\nn\\
\frac{\delta H}{\delta u}&\equiv& -\pi_u\,_t=\frac{\partial\mathcal{
H}}{\partial u}-\frac{\partial}{\partial x}\frac{\partial \mathcal{H}}{\partial u_x}=
\theta(-x)(-u_{xx}+U^{'})+\delta(x)\left(\mathcal{D}_u-\frac{\mu_2}{2}+u_x\right),\nn\\
\frac{\delta H}{\delta v}&\equiv& -\pi_v\,_t=\frac{\partial\mathcal{
H}}{\partial v}-\frac{\partial}{\partial x}\frac{\partial \mathcal{H}}{\partial v_x}=
\theta(x)(-v_{xx}+V^{'})+\delta(x)\left(\mathcal{D}_v+\frac{\mu_1}{2}-v_x\right),
\end{eqnarray}
where $U^{'}=U_u$ and $V^{'}=V_v$, the Poisson brackets \eqref{constraintsintime} can be calculated.
In consequence, \eqref{constraintsintime} leads to explicit expressions
for the functions $\mu_1$ and $\mu_2$, which are
$$\mu_1=-\mathcal{D}_v+v_x\quad \mu_2=\mathcal{D}_u+u_x.$$
Assembling all these ingredients, the Hamiltonian density \eqref{Hd} becomes
\begin{eqnarray}\label{Hdensity}
\mathcal{H}&=&\theta(-x)\left(\frac{\pi_u^2+u_x^2}{2}+U\right)+
\theta(x)\left(\frac{\pi_v^2+v_x^2}{2}+U\right)\nn\\
&&\ \ +\,\delta(x)\left[\left(\pi_u+\frac{v}{2}\right)(v_x-\mathcal{D}_v)+\left(\pi_v-
\frac{u}{2}\right)(u_x+\mathcal{D}_u) + \mathcal{D}\right].
\end{eqnarray}
Expressions \eqref{HCanEqu} are the canonical Hamilton equations and
using the definitions of the canonical momenta they
coincide with the defect conditions and equations of motion of the
type I defect problem (the latter by performing a differentiation with respect
to time).

\p In principle, the conservation of any charge can be verified by calculating its Poisson bracket with the Hamiltonian.  For example, consider the total momentum of the system, which is defined by
\begin{equation}\label{consm}
P=\int_{-\infty}^{\infty} dx\, \mathcal{P}\quad \mbox{with}\quad
\mathcal{P}=\theta(-x)\pi_u \,u_x+\theta(x)\pi_v\, v_x+\delta(x)\Omega(u,v).
\end{equation}
It is straightforward to calculate the time derivative of $P$ using its Poisson bracket with the Hamiltonian  to obtain,
\begin{eqnarray}
\dot{P}&=&\delta(x)\left[\frac{1}{2}\left(\D_u^2-\D_v^2\right)-U+V+u_t(\Omega_u-\D_v)+v_t(\Omega_v-\D_u)\right]=0.
\end{eqnarray}
The final step follows from the facts that $(\D_u^2-\D_v^2)/2=(U-V)$, $\D=(f+g)$ and $\Omega=(f-g)$, with
$f=f(p)$ and $g=g(q)$, as was described previously in \cite{bczlandau}.

\p It should be noticed that the constraints \eqref{Hconstraints} are second class. Hence, as mentioned at the beginning of this section, they indicate  that not all degrees of freedom are independent. By definition, a constraint is first class if its Poisson brackets with all other constraints are zero - the constraints themselves can be imposed, if needed - otherwise, it is second class. In the present case, it is straightforward to check that the Poisson brackets of the constraints are constant. In fact,
\begin{equation}
C_{ij}\equiv\{\chi_i,\chi_j\},\quad C=\left(
                                   \begin{array}{cc}
                                     \phantom{-}0 & \phantom{-}1 \\
                                     -1 & \phantom{-}0 \\
                                   \end{array}
                                 \right).\nn
\end{equation}
The matrix $C$ can be used to construct the Dirac brackets,  the standard tool for dealing with second class constraints.

\p Next, consider the type II defect and suppose the Lagrangian density is given by
\eqref{lambdalagrangian}.
Then, there are three fields $u$, $v$ and $\lambda$, whose canonical
momenta are
\begin{equation}
\pi_u=\frac{\partial \mathcal{L}}{\partial
u_t}=\theta(-x)\,u_t-\delta(x)\,\left(\frac{v}{2}-\lambda\right),\
\pi_v=\frac{\partial \mathcal{L}}{\partial
v_t}=\theta(x)\,v_t+\delta(x)\,\left(\frac{u}{2}-\lambda\right),\nn
\end{equation}
 \begin{equation}
\pi_\lambda=\frac{\partial \mathcal{L}}{\partial \lambda_t}=-\delta(x)(u-v).\nn
\end{equation}
Consequently, the primary constraints are
\begin{equation}
\chi_1=\pi_u+\frac{v}{2}-\lambda=0,\quad
\chi_2=\pi_v-\frac{u}{2}+\lambda=0,\quad \chi_3=\pi_\lambda+(u-v)=0,
\end{equation}
and the Hamiltonian density reads
\begin{equation}\label{Hdlambda}
\mathcal{H}=\theta(-x)\left(\frac{\pi_u^2+u_x^2}{2}+U\right)+
\theta(x)\left(\frac{\pi_v^2+v_x^2}{2}+V\right)+\delta(x)(\mathcal{D}+\mu_1\chi_1+\mu_2\chi_2+\mu_3\chi_3).
\end{equation}
Since these constraints must be consistent with the evolution equations,
their time derivative must vanish. By using the following Poisson
bracket
\begin{eqnarray}\label{PBlambda}
\{F,G\}&=&\int_{-\infty}^{\infty}\,dx\left(\frac{\delta F}{\delta
u}\frac{\delta G}{\delta \pi_u}-\frac{\delta F}{\delta
\pi_u}\frac{\delta G}{\delta u}\right)+\int_{-\infty}^\infty
\,dx\left(\frac{\delta F}{\delta v}\frac{\delta G}{\delta
\pi_v}-\frac{\delta F}{\delta \pi_v}\frac{\delta G}{\delta v}\right)\nn\\
&&\hskip 2.5cm  +\left(\frac{\delta F}{\delta \lambda}\frac{\delta G}{\delta
\pi_\lambda}-\frac{\delta F}{\delta \pi_\lambda}\frac{\delta G}{\delta
\lambda}\right)_{x=0},
\end{eqnarray}
it is possible to verify that
\begin{equation}
\chi_1\, _t=-\mathcal{D}_u-u_x+\mu_2-2\mu_3=0,\
\chi_2\, _t=-\mathcal{D}_v+v_x-\mu_1+2\mu_3=0,\
\chi_3\, _t=-\mathcal{D}_\lambda+2(\mu_1-\mu_2)=0.\nn
\end{equation}
Unlike the previous case, this system of equations does not
determine completely the functions $\mu_j$. In fact, requiring the constraints
to be preserved with time
forces
\begin{eqnarray}
\mu_1&=&-\mathcal{D}_v+ v_x+2\mu_3,\quad \mu_2=\mathcal{D}_u+
u_x+2\mu_3\label{priC}\\
&&(u-v)_x+\mathcal{D}_u+\mathcal{D}_v+\frac{1}{2}\mathcal{D}_\lambda=0\label{secC}.
\end{eqnarray}
Expression \eqref{secC} is a secondary constraint. However, it
 is not genuinely new
since it coincides with an algebraic sum of some of the canonical Hamiltonian equations, as
can be verified by using the following Hamiltonian density
\begin{eqnarray}\label{Hdensitylambda}
\mathcal{H}&=&\theta(-x)\left(\frac{\pi_u^2+u_x^2}{2}+U\right)+
\theta(x)\left(\frac{\pi_v^2+v_x^2}{2}+V\right)+\delta(x)\mathcal{D}\nn\\
&+&\delta(x)\left[\left(\pi_u+\frac{v}{2}-\lambda\right)(v_x-\mathcal{D}_v)+
\left(\pi_v-\frac{u}{2}+\lambda\right)(u_x+\mathcal{D}_u)+\mu_3(2\pi_u+2\pi_v+\pi_\lambda)\right].\nn\\
\end{eqnarray}
In fact,
$$0=(\pi_\lambda+2\pi_u+2\pi_v)\, _t\equiv -(u-v)_x-\mathcal{D}_u-\mathcal{D}_v-\frac{1}{2}\mathcal{D}_\lambda,$$
which coincides with \eqref{secC}.
As was shown in the previous case, all Hamilton equations can be obtained and they lead to the
equations of motion and defect conditions (note that $\lambda_t\equiv\mu_3$). Finally, as
mentioned before, the Poisson brackets \eqref{PBlambda} may be used to
verify the conservation of charges. For example, given the total momentum
\eqref{consm}, it can be checked that
$$
\dot{P}=\delta(x)\left[\frac{1}{2}\left(\D_u^2-\D_v^2\right)-U+V+\lambda_t(\mathcal{D}_\lambda+2\mathcal{D}_u+2\mathcal{D}_v+\Omega_\lambda)
+u_t(\Omega_u-D_v)+v_t(\Omega_v-D_u)\right].\nn
$$
Since $D=(f+g)$ and $\Omega=(f-g)$, with $f=f(p-\lambda,q)$ and
$g=g(\lambda,q)$, the above expression  becomes
$$
\dot{P}=\delta(x)\left[\frac{1}{2}\left(\D_u^2-\D_v^2\right)-U+V+\frac{1}{2}f_q\mathcal{D}_\lambda\right]\equiv 0.
$$
In summary, from the Hamiltonian density \eqref{Hdensitylambda}, it is possible to read off the final constraints, which are
$$\chi_1,\quad \chi_2,\quad \gamma_1=2\pi_u+2\pi_v+\pi_\lambda,$$
where $\chi_1$, $\chi_2$ are second class, while $\gamma_1$ is first class. In fact, it can be checked that $\{\chi_1, \gamma_1\}=\{\chi_2, \gamma_1\}=\{\gamma_1, \gamma_1\}=0$. The first class constraints are usually related to the presence of a gauge freedom. In the type II defect framework, the existence of a first class constraint indicates the freedom to translate the field $\lambda$ by any function of $q$, as was pointed out in section 3.

\section{Comments and conclusions}

\p The main result of this paper has been to extend the framework within which an integrable defect may be described. The previous framework (referred to as type I in this article) seemed fairly natural yet even for a single scalar field was unable to accommodate all possible relativistic integrable models because the Tzitz\'eica, or $a_2^{(2)}$ affine Toda, model was conspicuously absent. For multiple scalar fields the possible type I defects are restricted to the $a_n^{(1)}$ series of affine Toda models. In all cases, the type I defects are intimately related to B\"acklund transformations, in the sense that the conditions relating the fields on either side of an integrable defect take the form of a B\"acklund transformation frozen at the location of the defect. At first sight, this relationship seemed attractive since it provided a use for B\"acklund transformations that had not been noticed before. On the other hand, the Tzitz\'eica equation has several B\"acklund transformations associated with it and none of them emerged naturally from within the type I framework. Moreover, the integrability of the type I defects is intimately related to momentum conservation, in the sense that insisting there should be a total momentum including a contribution from the defect itself leads to restrictions that would be associated normally with the requirements of having higher spin conserved quantities. It is a curious situation: certain integrable systems (those with type I defects) can violate translational invariance yet preserve momentum. The question is: can this phenomenon be extended to other integrable systems by changing the framework? It appears the answer is yes, and one particular different framework (referred to as type II) is described in this paper. In fact,  only a slight change appears to be necessary, the Tzitz\'eica model is incorporated, and the relationship with frozen B\"acklund transformations is modified. The trick is to introduce a new degree of freedom located on the defect and couple it in a minimal manner to the discontinuity across the defect. In the absence of a generalised Lax pair for the type II system, momentum conservation becomes a tool for identifying the possibilities, backed up by other less direct evidence. Turning the argument around and starting from the defect conditions allows an apparently new B\"acklund transformation to be established for the Tzitz\'eica equation. The type II framework certainly contain all single field integrable systems of Toda type (or free fields) but it is not yet demonstrated these are the only possibilities. The latter appears reasonable since \eqref{fgrelation} is highly constraining but a complete proof of integrability needs to be found in order to be sure.

\p It is already known that the $a_n^{(1)}$ affine Toda models can support type I defects of several kinds and that defects are able to relate different $a_n$ conformal Toda models to each other (thereby generalising the relationship between the Liouville model and free fields \cite{cz09}). However, other affine Toda models based on the root data of the $b,c,d,e,f,g$ series of Lie algebras do not appear to fit in to the type I framework. This is surprising:  in most respects, the affine Toda field theories at least in the bulk, have similar features, though it does appear from the literature that the $a_n^{(1)}$ series is special in having a B\"acklund transformation of a simple type. It remains to be seen if the type II framework can be adapted to all Toda models. The folding process cannot explain the apparent difficulties with the $d,e$ series. However, once these are understood the folding process might be an essential part of the story for the remaining cases. For that reason it would be natural to examine the $d,e$ series next.

\p At this stage it is worth outlining a possible direction for a generalisation containing multi-component fields. Using the same notation as previously, taking as a starting point the defect contribution
\begin{equation}\label{multicomponentdefectlagrangian}
 {\cal L}_D=\delta(x)\left(q\cdot A q_t + 2 \lambda\cdot q_t -{\cal D}(\lambda,q,p)\right),
\end{equation}
where $A$ is an antisymmetric matrix, then insisting on overall momentum conservation, leads to the following
constraints on ${\cal D}$ and $\Omega$:
\begin{equation}\label{DandOmega}
{\cal D}=f(p-\lambda,q) + g(p+\lambda,q),\quad \Omega = f(p-\lambda,q) - g(p+\lambda,q).
\end{equation}
Further, the two functions $f$ and $g$ are constrained by a generalisation of the Poisson bracket relation
\eqref{fgrelation} that reads,
\begin{equation}\label{generalfgrelation}
 \nabla_q f\cdot\nabla_\lambda g - \nabla_q g\cdot\nabla_\lambda f +\nabla_\lambda f\cdot A\, \nabla_\lambda g = U(u) - V(v).
\end{equation}
Here $A$ is the antisymmetric matrix occurring in \eqref{multicomponentdefectlagrangian} and $U,\ V$ are the bulk potentials for the fields to either side of the defect. The left hand side of \eqref{generalfgrelation} is a bona fide Poisson bracket since it is antisymmetric and satisfies the Jacobi relation, yet, as before, all dependence on $\lambda$ must cancel out. This provides severe constraints on $U$ and $V$, which will be explored elsewhere.

\p At the quantum level, it was demonstrated in \cite{bczsg05, cz07} that type I defects within the $a_n^{(1)}$ series are described by infinite-dimensional transmission matrices, which are determined up to a single parameter by a set of `triangle relations' ensuring their compatibility with the bulk S-matrix. Moreover, arguments have been provided to demonstrate that the free parameter is essentially the same, though possibly renormalised, as the free parameter in the type I Lagrangian. Clearly, the next question concerns the transmission matrix in the context of type II defects. For the sine-Gordon model, the transmission matrix in this framework should depend on two independent parameters and there should be some evidence or influence of the confined field $\lambda$, at least recognising the $i\pi$ ambiguity mentioned in section 4.1.  At a quantum level, the Tzitz\'eica model contains a triplet of equal mass states, reflecting its origin in $a_2^{(1)}$ affine Toda field theory under the folding process, only two of which correspond to classical solitons, and its S-matrix is known \cite{smirnov}.  It is to be hoped there will be a transmission matrix based on an ansatz that takes into account the mysterious role of $\lambda$ (this time the ambiguity is threefold - see section 4.2).

\bigskip\p{\bf Acknowledgements}

\bigskip

\p We are grateful for conversations with colleagues in Durham, especially Peter Bowcock. In particular, we wish to thank him for discussions on the content of section 5, much of which he developed independently.

\p We also wish to express our gratitude to the UK Engineering and Physical Sciences
Research Council for its support under grant reference EP/F026498/1.

\appendix
\section{Energy-like spin three charge for the sine-Gordon model}
\label{appendixA}

\p In this appendix it is shown that an energy-like spin three charge for the sine-Gordon model with a defect of type II is conserved. The bulk charge, which is not expected to be conserved in the presence of a defect, conveniently normalised, reads
\begin{eqnarray}
{\cal E}_3&=&\int^{0}_{-\infty}dx \,\left(\frac{u_t^4+u_x^4}{4}+\frac{3}{2}u_x^2 u_t^2+4 u^2_{tx}+(u_{tt}+u_{xx})^2
+(u_t^2+u_x^2)U^{''}\right) \nn\\
&&\ \ +\int^{\infty}_{0}dx \,\left(\frac{v_t^4+v_x^4}{4}+\frac{3}{2} v_x^2 v_t^2+4 v^2_{tx}+(v_{tt}+v_{xx})^2
+(v_t^2+v_x^2)V^{''}\right),\nn
\end{eqnarray}
and its time derivative is

\begin{eqnarray}\label{tdspin3d}
\dot{\cal E}_3
&=&\left[(u_tu_x^3+u_t^3u_x)-(v_tv_x^3+v_t^3v_x)
 +4(2u_{tt}+U^{'})u_{tx}\right.\nn\\
 &&\ \ \ \ \ \ \ \ \ \ \ \ \ \ \ \ \ \ \ \left.-4(2v_{tt}+V^{'})v_{tx}-2(u_tu_xU^{''}-v_tv_xV^{''})\right]_{x=0},
\end{eqnarray}
where $U^{'}=U_u$ and $V^{'}=V_v$. This is not expected to be zero but the right hand side may turn out to be the total time derivative of a functional $-\D_3$ that depends only on the defect variables $p,\ q$ and $\lambda$. In that case, ${\cal E}_3+\D_3$ will be conserved. Since the expression \eqref{tdspin3d} is calculated at $x=0$, it is convenient to rewrite it
by using the variables $p$ and $q$. Then,  using the defect conditions \eqref{defectcondition}-\eqref{defectcondition3} with the functions $f$ and $g$ given by \eqref{sgfandg}, the expression \eqref{tdspin3d} becomes a total time derivative
\begin{eqnarray}\label{spin3charge}
\dot{\cal E}_3
&=&4\frac{d}{dt}\left(2(p_t-\lambda_t)q_tf_{\lambda q}-(p_t-\lambda_t)^2f-q_t{^2}(f+g)_{qq}-\lambda_t{^2}g
-2q_t\lambda_tg_{\lambda q}\right)\nn\\
&&+4\frac{d}{dt}\left((p_t-\lambda_t)(U^{'}-V^{'})-\lambda_t(U^{'}-V^{'})-q_t(U^{'}+V^{'})\right)-\frac{d}{dt}\Omega_3(p,q,\lambda),
\end{eqnarray}
(where again on the right hand side all field quantities are evaluated at $x=0$), with
\begin{eqnarray}
\frac{\partial\Omega_3}{\partial q}&=&3f_\lambda(U-V)-\frac{3}{4}f_q(f+g)_\lambda{^2}
+\frac{1}{4}(f+g)_q\left(3f_\lambda{^2}+(f+g)_q{^2}-12(U+V)\right),\nn\\
\frac{\partial\Omega_3}{\partial p}&=&-3(f+g)_q(U-V)+\frac{3}{4}f_{qq}(f+g)_\lambda{^2}-\frac{1}{4}f_\lambda
\left(f_\lambda{^2}+3(f+g)_q{^2}-12(U+V)\right),\nn\\
\frac{\partial\Omega_3}{\partial \lambda}&=&\frac{1}{4}(f+g)_\lambda\left(f_\lambda{^2}+g_\lambda{^2}-f_\lambda g_\lambda-
3(f+g)_\lambda(f+g)_{qq}+3(f+g)_q{^2}-12(U+V)\right).\nn
\end{eqnarray}
The formula \eqref{spin3charge} has been obtain by making use of the following properties of the defect potential for the sine-Gordon model
\begin{equation}\label{helpformulae}
f_p=-f_\lambda,\quad f_{\lambda\lambda}=f, \quad g_{\lambda\lambda}=g, \quad f_{qqq}=f_q,
\quad g_{qqq}=g_q,\quad f_{\lambda q}=f_q, \quad g_{\lambda q}=g_q.
\end{equation}
Finally, it has been verified that the cross derivatives of the function $\Omega_3$ are consistent, that is
$$\frac{\partial^2\Omega_3}{\partial q\partial p}=\frac{\partial^2\Omega_3}{\partial p\partial q},\quad
\frac{\partial^2\Omega_3}{\partial q\partial \lambda}=\frac{\partial^2\Omega_3}{\partial \lambda\partial q},\quad
\frac{\partial^2\Omega_3}{\partial p\partial \lambda}=\frac{\partial^2\Omega_3}{\partial \lambda\partial p}.$$
For this task, in addition to \eqref{helpformulae},  the following relations have been used
\begin{equation}
(U\pm V)_p=(U\mp V)_q, \quad f_{qq}(f+g)_q=f_q(f+g)_{qq},\quad f_q(g+g_\lambda)=g_q(f+f_\lambda)
\end{equation}
where
$$(U-V)=\frac{1}{2}(f_q g_\lambda-f_\lambda g_q)=(U+V)_{pq}.$$

\end{document}